\documentclass[prl,nofootinbib,preprintnumbers,tightenlines,twocolumn,superscriptaddress]{revtex4-1}
\usepackage{braket}
\usepackage{array}
\usepackage{dsfont}
\usepackage{amsmath}
\usepackage{pifont}
\usepackage[normalem]{ulem}
\usepackage{amssymb}
\usepackage{xcolor}
\usepackage{amsfonts}
\usepackage{mathtools}
\usepackage{graphicx}
\usepackage{dcolumn}
\usepackage{bm}
\usepackage{multirow}
\usepackage{hhline}
\usepackage{siunitx}

\usepackage{leftidx}
\usepackage{color}
\usepackage{mathtools}
\usepackage{MnSymbol}
\usepackage[mathscr]{eucal}
\usepackage[german,english]{babel}
\usepackage[capitalize]{cleveref}

\begin{document}
\title{Transitional supersolidity in ion doped helium droplets}
\author{Juan Carlos Acosta Matos}
\affiliation{Max-Planck-Institut f\"ur Physik komplexer Systeme, N\"othnitzer Str.\ 38, D-01187 Dresden, Germany }

\author{P. Giannakeas}
\affiliation{Max-Planck-Institut f\"ur Physik komplexer Systeme, N\"othnitzer Str.\ 38, D-01187 Dresden, Germany }
\author{Matteo Ciardi}
\affiliation{Institute for Theoretical Physics, Vienna University of Technology, Wiedner Hauptstraße 8-10/136, A-1040 Vienna, Austria}
\author{Thomas Pohl}
\affiliation{Institute for Theoretical Physics, Vienna University of Technology, Wiedner Hauptstraße 8-10/136, A-1040 Vienna, Austria}
\author{Jan-Micheal Rost}
\affiliation{Max-Planck-Institut f\"ur Physik komplexer Systeme, N\"othnitzer Str.\ 38, D-01187 Dresden, Germany }

\begin{abstract}
\noindent
$^4$He nanodroplets doped with an alkali ion feature a snowball of crystallized layers surrounded by superfluid helium.   
For large droplets, we predict that a transitional supersolid layer can form, bridging between the solid core and the liquid bulk, where the $^4$He density displays modulations  of icosahedral group symmetry.
To identify the different phases, we combine density functional theory with the semiclassical Gaussian time-dependent Hartree method for localized many-body systems. This hybrid approach can handle large particle numbers and provides insight into the physical origin of the supersolid layer.
For small droplets, we verify that the predictions of our approach are in excellent agreement with Path-Integral Monte Carlo calculations. 
%Our analysis demonstrates that locally the supersolid layer possesses a finite non-classical translational and rotational inertia even at finite temperature showcasing its stability on thermal fluctuations.
\end{abstract}
\maketitle

Supersolidity is a counterintuitive aspect of quantum matter, featuring the apparently incompatible properties of rigid crystalline order and superfluid flow due to translational and global gauge symmetry breaking \cite{grossPR1957}. Owing to its universal character, supersolidity can occur in very different physical systems, e.g., in Bose-Einstein condensates with spin-orbit coupling \cite{li2017stripe}, in cavities \cite{leonard2017supersolid}, or in dipolar gases \cite{tanziprl2019, chomazprx2019,boettcherPRX2019}. 
In the latter case, recent experiments demonstrate that supersolidity arises as strong density modulations, where the system fragments into connected clusters of superfluid droplets \cite{tanziprl2019, chomazprx2019,boettcherPRX2019,boettcher_2021,norcia2021two,biagioniprx2022}.
The possibility of fine-tuning the interactions to tailor the properties of dipolar gases has motivated broad theoretical studies of supersolids in three-dimensional \cite{ancilottopra2013, bombinprl2017, zhangprl2019,zhangpra2021} and low-dimensional set-ups \cite{cintipra2017, wenzelpra2017,baillieprl2018} or in bubble traps \cite{ciardiprl2024}, for a recent review see \cite{rest23}.
 In all these examples supersolidity arises from the emergence of a density wave triggered by a roton instability in the excitation spectrum of the superfluid, which spontaneously breaks the continuous translational symmetry of the system, as  originally proposed for bulk helium \cite{grossPR1957,GROSS195857}.

In view of the strong atomic interactions in solid helium, another concept of supersolidity was suggested \cite{andreev1969quantum,chesterPRA1970,boninsegnirpm2012}. This so called Andreev-Lifshitz supersolidity (ALS) is induced by mobile defects which may form in the groundstate of the crystal and induce a small but finite superfluid response. ALS has also been proposed theoretically for helium droplets doped with molecular structures of distinct symmetry, such as C$_{20}$ \cite{kwonprb2010} or CH$_5^+$ 
\cite{brieucManifestationsLocalSupersolidity2023}.
Despite continuing efforts, there is not yet consensus on whether ALS was observed in experiments \cite{chan2013overview}.

\begin{figure}[t!]
\centering
\includegraphics[scale=0.41]{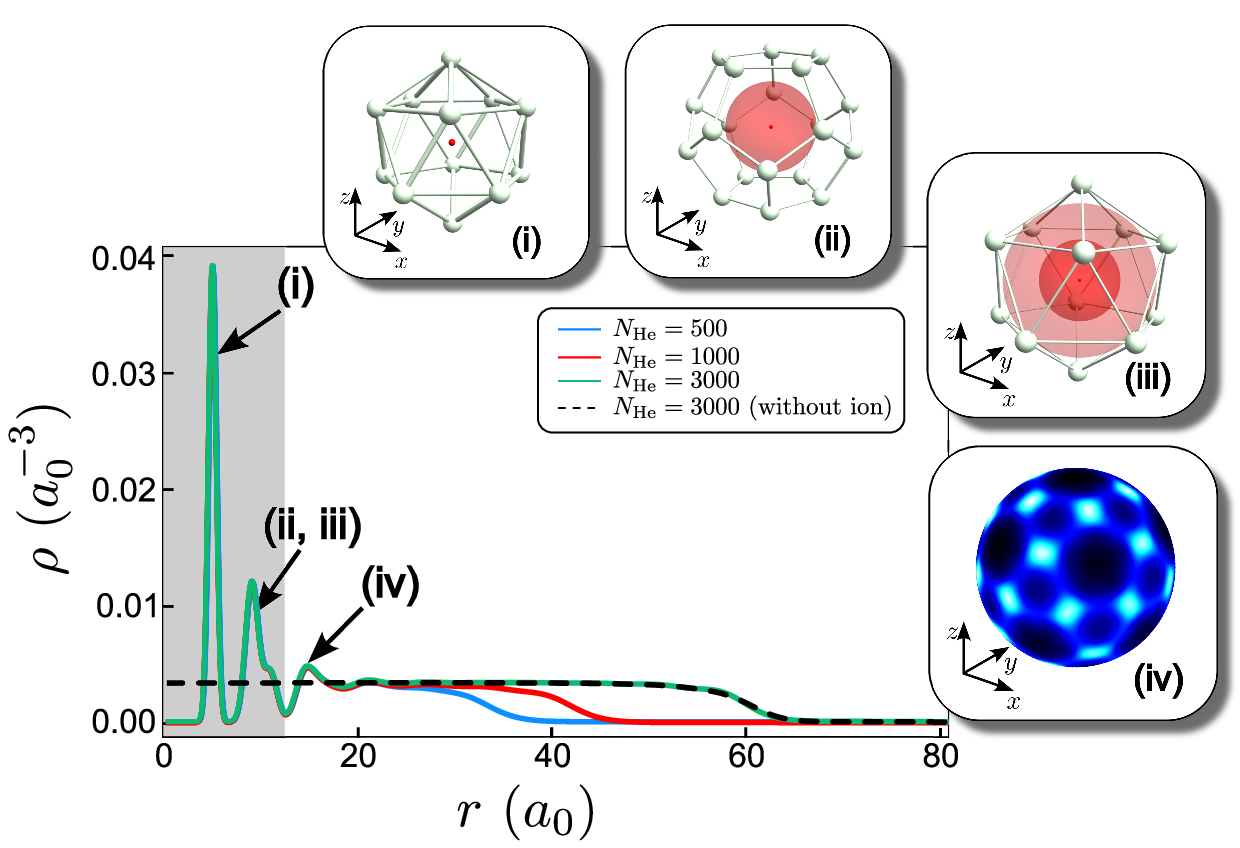}
\caption{The radial density of a He droplet doped with $\rm{Na}^+$ for different sizes $N_{\rm{He}}= 500~\rm{(blue)}$, $1000~\rm{(red)}$, $3000~\rm{(green)}$  and for the
prestine droplet with $N_{\rm{He}}= 3000$ (dashed). The gray shaded area denotes the radial region of snowball formation. This region exhibits two crystallized layers and the corresponding He cluster configuration is shown in Panels (i)-(iii), where red spheres indicate the size of the next smaller layer.
Panel (iv) shows the density  in the third, transitional layer, where the He atoms exhibit supersolid behavior.}
\label{fig:fig1}
\end{figure}
Here, we demonstrate that a $^4$He droplet can give rise to a mesoscopic density modulation-induced supersolid (DMS) if doped with a suitable ion. This establishes DMS for a finite He system, where the symmetry-breaking mechanism is due to the polarization forces of the ion in the center of the droplet, and not due a roton instability, as in ultracold gases.

The isotropic ion-He interactions modulate the droplet's density radially, by forming layers (here and in the following He stands for $^4$He). They can give rise to solvation shells, on which the He atoms organize themselves in a crystal structure, the so-called {\it snowball}, see Fig. 1, (i)-(iii). The snowball effects have been extensively studied theoretically \cite{buzzacchiAlkaliIonsSuperfluid2001, rossiAlkaliAlkaliearthIons2004,cocciaBosonicHeliumDroplets2007,chikinaPRB2007,zunzunegui-bruACS2023a,bartlJPCA2014,slavicekPCCP2010,slavicekSnowballsQuantumSolvation2010,tramontoPathIntegralMonte2015,rastogiLithiumIonsSolvated2018} and experimentally \cite{popitrenaudprl1972,marinettiMicrosolvationCationicDimers2007,muller2009alkali,bartl2014size,albrechtsen2023observing}.

In the outer region, the polarizing ion-He interaction is weak, leaving the atoms in a superfluid phase. 
For large nanodroplets, a transitional layer can form (Fig.~\ref{fig:fig1}, (iv)) whose  density is structured by the interaction with the He snowball. This structure can give rise to DMS, as we will discuss below.

Our exemplary system giving rise to DMS  consists of a He nanodroplet doped with a $\rm{Na}^+$ ion described by
\begin{align}
    \hat{H}&=\sum_{n=1}^N\frac{\hat{P}^2_n}{2 M_\mathrm{He}}  +\sum_{m<n}\hat{V}_{nm}+\frac{\hat{P}_I^2}{2 M_I}+\sum_{n=1}^N \hat{V}_{In},
    \label{eq:eq1}
\end{align}
where $M_\mathrm{He}$ ($M_I$) is the mass of He ($\rm{Na}^+$) atoms (ion), and $\hat{V}_{nm}$ [$\hat{V}_{In}$] refers to the $\rm{He}-\rm{He}$ [$\rm{Na}^+-\rm{He}$] interaction.
The Hamiltonian  \cref{eq:eq1}, with or without the presence of an ion, can be efficiently tackled within the framework of the Orsay-Trento density functional theory (OT-DFT) \cite{dalfovoStructuralDynamicalProperties1995}. This method has been extensively used, in particular over the last decade, to capture physics in the limit of large system sizes
\cite{lehtovaaraEfficientNumericalMethod2004,ancilottoDensityFunctionalTheory2017,longDensityFunctionalTheory2021}.

For a pristine He nanodroplet, OT-DFT predicts a radially flat profile of the He density, reflecting the incompressibility of the liquid droplet (black line in \cref{fig:fig1}).
In contrast, with a $\rm{Na}^+$ ion in the droplet's center, the radial density exhibits pronounced peaks (green line in \cref{fig:fig1}) leading to the snowball regime (gray shaded area in \cref{fig:fig1}) \cite{atkinsIonsLiquidHelium1959}, in which almost all He atoms crystallize on the inner two layers. This happens because the ion induces polarization effects \cite{cocciaBosonicHeliumDroplets2007}, pulling He atoms towards itself.

On a mesoscopic level, OT-DFT quantitatively predicts the radial density,
location of the shells, the size of the snowball, and its collective excitation spectrum, even for large, ion-doped droplets.
However, this approach does not provide microscopic insight, e.g., the particle configuration of the snowball, or the He angular density distribution in the transitional layer (iv) in \cref{fig:fig1}.
Such properties typically require many-body methods, e.g., multiconfiguration time-dependent Hartree \cite{meyerMulticonfigurationalTimedependentHartree1990, caoUnifiedInitioApproach2017}, or Monte Carlo based methods \cite{buzzacchiAlkaliIonsSuperfluid2001,rossiAlkaliAlkaliearthIons2004,cocciaBosonicHeliumDroplets2007,galliPathIntegralMonte2011,tramontoPathIntegralMonte2015}, which are restricted to small particle numbers.  

To describe larger droplets, while retaining information about their microscopic structure, we employ a semiclassical approach. 
It relies on the fact that the $N_s$ He atoms contributing to the snowball structure are very well localized with negligible particle exchange in sharp contrast to a superfluid.
Furthermore, since the snowball (gray region in \cref{fig:fig1}) is nearly spatially separated from the superfluid part of the droplet, we omit the exchange between the $N_s$ atoms of the snowball and the $N_d=N_{\rm{He}}-N_s$ particles which comprise the residual droplet. 
Hence, for the snowball region, we use the G-TDH method of \cite{unn-tocQuantumDynamicsSolid2012}, where the
resulting equations of motion consist of a classical ensemble of trajectories capturing the main quantum effects of localized particles. 
The $N_s$ atoms are represented by a direct product of single particle Gaussians, i.e., $\ket{\boldsymbol{z}_n(\tau)} \equiv \ket{\boldsymbol{r}_{0n}(\tau), \alpha_n(\tau)}$, where the parameter $\boldsymbol{r}_{0,n}(\tau)$ [$\alpha_n(\tau)$] relates to the centroid (width) of the $n$-th Gaussian at an instance $\tau$.
These parameters are obtained by evolving the subset of $N_s$  atoms in imaginary time until it equilibrates [for details see A in End Matter].
Within the G-TDH approach, the first layer [ \cref{fig:fig1} (i)] exhibits an icosahedron symmetry of a radius $\sim 4.88~\rm{a_0}$.
Furthermore, contrary to OT-DFT, the G-TDH method predicts that the second layer [panels (ii) and (iii)] consists of two subshells of dodecahedron and icosahedron symmetry, with radii $\sim 9.19~\rm{a_0}$ and $\sim 11.45~\rm{a_0}$, respectively.
The predicted shell substructure is in excellent agreement with the Path-Integral Monte Carlo (PIMC) calculations at finite temperature \cite{galliPathIntegralMonte2011}.

\begin{figure}[t!]
\centering
\includegraphics[scale=0.85]{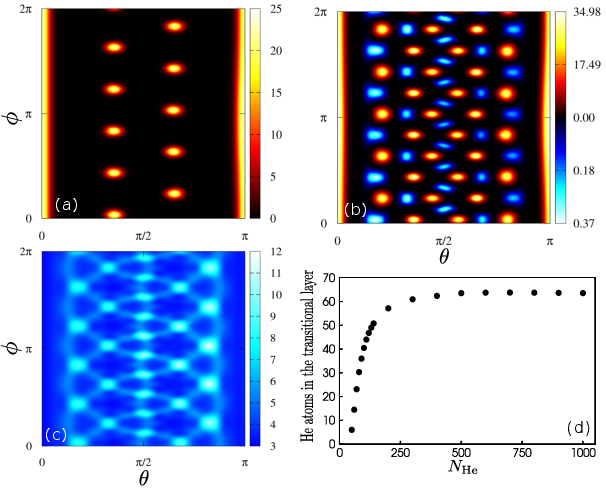}
\caption{Angular density distribution for the first (a), second (b) and third (c) layer for a droplet of $N_{\rm{He}}=1000$. The distributions are radially averaged between two successive minima of the density shown in \cref{fig:fig1}. The black-to-yellow (black-to-blue) color gradient indicates the density of  He atoms obtained via G-TDH (mDFT).
Panel (d) shows the number of atoms that occupy the transitional layer as a function of the total number  $N_{\rm{He}}$ of atoms.}
\label{fig:fig2}
\end{figure}

Beyond the snowball region, the $N_d$ atoms of the residual droplet are delocalized and particle exchange dominates, rendering the G-TDH approach invalid.
Therefore, we develop a time-dependent modified density functional theory (mDFT) framework that describes the rest of the droplet.
More specifically, the standard Orsay-Trento total energy functional is utilized with the snowball region included as an external potential $U_S(\boldsymbol{r})$ emerging from G-TDH. In this way,
mDFT retains the short-range correlation effects whose impact on the superfluid part of the ion-doped droplet is crucial.
Formally, the modified energy functional obeys the relation $E[\Psi,\Psi^*]=E_{OT}[\Psi,\Psi^*]+V_{\rm{ext}}[\rho]$, where $E_{OT}[\Psi,\Psi^*]$ denotes the Orsay-Trento functional [for details see B in End Matter].
The external potential $V_{\rm{ext}}[\rho]$ is given by
\begin{subequations}
\begin{align}
    V_{\rm{ext}}[\rho] &= \int  d^3 \mathbf{r} \rho(\mathbf{r})  \left[ V_I(r) + U_S(\mathbf{r})   \right], ~~\rm{with}  \label{eq:DFT_ExternalPotential}\\
    U_S(\mathbf{r}) &= \sum_{n=1}^{N_s} \int d \boldsymbol{r}_n |\braket{\boldsymbol{r}_n|\boldsymbol{z}_n(\tau \to \infty)}|^2 V(|\mathbf{r}-\mathbf{r}_n|).
\end{align}
\end{subequations}

With mDFT, we obtain the angular distribution of the reduced one-body density of the transitional layer [see \cref{fig:fig1} (iv)], which displays supersolid behavior: the He density maintains an ordered, solid-like configuration, although it is dominated by particle exchanges due to the droplet's superfluidity. 
%Hence, the system exhibits signatures of a solid and a superfluid simultaneously.
This behavior is enabled by the location of the He atoms on the spherical transitional layer, which ensures that all of them experience the same He-ion attraction. At the radius of the transition layer, its strength is comparable to that of the He-He interaction determined by the superfluid He density.
At radii beyond the transitional layer, the He-ion interaction is significantly weaker than the He-He interactions, resulting in a uniform angular distribution of the reduced one-body density as expected for a liquid droplet.

The difference between solid (snowball) layers and the supersolid transitional one can be further analyzed by examining the spatial arrangement of atoms in \cref{fig:fig2}.
For large droplets with $N_{\rm{He}}=1000$, \cref{fig:fig2} (a)-(c) illustrate the angular distribution of the reduced one-body density for all three layers.
For each of them, the corresponding distribution is radially averaged between consecutive minima of the radial density (see \cref{fig:fig1}) where the black-to-yellow (black-to-blue) color gradient corresponds to the $N_s$ ($N_d$) atoms of the snowball (residual droplet) region.
In the snowball region, panels (a) and (b) display well-localized atoms (black-to-yellow color gradient) due to negligible exchange where the first layer is completed with 12 atoms in icosahedron symmetry and the second one consists of 32 atoms.

In the second layer (panel (b)) one sees in addition to localized He density also a small contribution from the residual droplet density (black-to-blue color gradient) originating from  delocalized but structured He density of the transitional layer.
This behavior is expected, since the radial density between the second and third layers is suppressed but not zero (see green line in \cref{fig:fig1}). 
Panel (c) shows that the angular distribution density of the third, transitional layer, exhibits modulations of icosidodecahedron symmetry, despite the fact that atoms are delocalized.

Finally, panel (d) shows, for a large droplet of $N_{\rm{He}}=1000$, that the number of atoms in the transitional layer exceeds the number of maxima of the reduced one-body density.
Moreover, panel (d) demonstrates that the transitional layer's He density saturates for droplet sizes $N_{\rm{He}}\ge250$.
For these droplet sizes, the transitional layer$-$despite the finite number of He atoms that participate$-$displays the main idiosyncrasies of a supersolid with a polyhedron symmetry, induced by the snowball structure. The layer's spherical surface ensures that, on the layer, all atoms have similar ion-atom interaction which in turn is comparable to the atom-atom interactions.
%As mentioned earlier, the supersolid behavior emerges due to comparable  ion-atom and atom-atom interactions together with the layer's spherical surface, while the particular polyhedron symmetry is induced by the snowball structure.
Indeed, we estimate that, on the transitional layer, the atom-ion potential is $V_I(\boldsymbol{r}_{\rm TL})\approx -5\cdot10^{-6}~\rm{a.u.}$, whereas the $\rm{He}-\rm{He}$ potential is $V(\boldsymbol{r}_{\rm{NN}})\approx -3\cdot10^{-6}~\rm{a.u.}$, having called $\boldsymbol{r}_{\rm{NN}}$ the distance between neighboring density maxima on the layer.
Although this comparison demonstrates that the interactions are of the same order, $V_I$ is slightly more attractive, which enables the formation of the transitional layer.
Similar estimates for the other two layers reveal that $V_I$ dominates by orders of magnitude, while in the liquid beyond the transitional layer the $\rm{He}-\rm{He}$ potential prevails.

In order to clarify that the supersolid layer is not an artifact of our G-TDH/mDFT approach which constructs the snowball region as an emergent external potential, we have carried out PIMC calculations for smaller droplet sizes. Moreover, we have determined the superfluid fraction with both, the hybrid G-TDH/mDFT and the PIMC approach.

\begin{figure}[t!]
\centering
\includegraphics[scale=0.5]{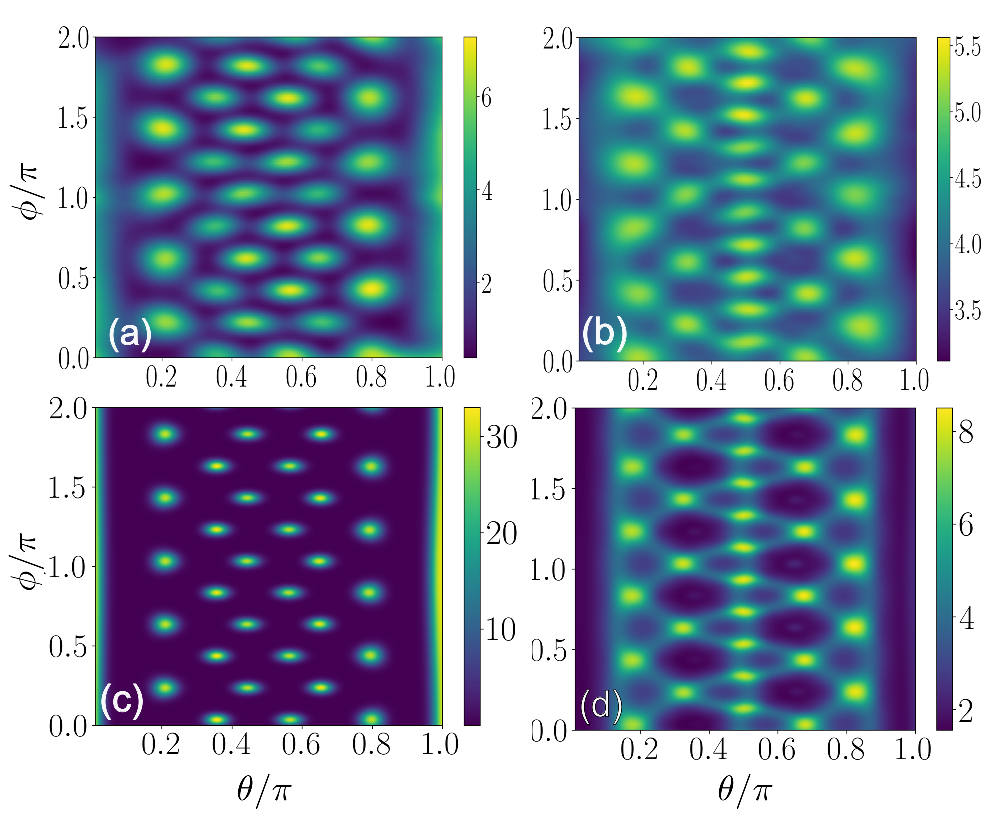}
\caption{Angular distribution of the reduced one-body density for $N_{\rm{He}}=128$. Panels (a) and (b) show PIMC calculations at $T=0.5~\rm{K}$ ($T=0~~\rm{K}$) for the second and third layer, respectively, and panels (c) and (d) show the result of the
G-TDH/mDFT calculation for the same layers. 
Note that density shown for the third layer starts from a finite value, as there is a background.
}
%Note that the G-TDH/mDFT results correspond to zero temperature and therefore  exhibit more localized maxima.}
\label{fig:fig3}
\end{figure}

PIMC is an \textit{ab initio} method for finite temperatures \cite{ceperley1995}, developed to provide accurate results for helium.
We have employed both, the worm algorithm to efficiently sample bosonic configurations \cite{boninsegni2006}, and the pair-product approximation based on the exact calculation of the two-body propagator at $T=20~\rm{K}$ or $T=40~\rm{K}$, described, e.g., in \cite{pollock1984}.  For droplet size $N_{\rm{He}}=128$, in \cref{fig:fig3}, we compare the angular distributions of the one-body density in the second and the transitional layer obtained from PIMC calculations at $T=0.5 ~~\rm{K}$ [panels (a) and (b)] with those from G-TDH/mDFT calculations at $T=0 ~~\rm{K}$ [panels (c) and (d)]. That the maxima are more localized in (c) and (d) is due to the fact that the  G-TDH/mDFT calculations have been carried out at zero temperature.
Both calculations show the polyhedron symmetry of the second shell, and the results are in excellent agreement, validating the G-TDH approach.

More importantly, PIMC predicts virtually the same density modulations as mDFT for the transitional layer, indicating a symmetry breaking in the rotational degrees of freedom.
This implies that our hybrid approach captures the physical origin of DMS correctly with the emergent external potential. The PIMC calculation confirms that, despite the isotropic ion-atom and atom-atom interactions, the clustering of few He atoms in form of the snowball triggers the pattern formation of DMS in the transitional layer.

As final evidence for DMS of the transitional layer, 
 we determine in both approaches the local superfluid fraction as a key quantity for identifying supersolid behavior by assessing to what extent the helium atoms move frictionlessly. 
 With mDFT, we compute the non-classical translational inertia (NCTI) akin to \cite{PomeauPRL1994,sepulveda2010superfluid,roccuzzoSupersolidBehaviorDipolar2019}.
The stationary order parameter $\Psi_0(\boldsymbol{r})$ obtained via mDFT is perturbed by applying a uniform velocity, e.g., $v_x$ along the $x$-axis. This means to include a {phase twist} in the  initial conditions for solving the time-dependent mDFT  in the lab frame.
The phase twist  $\Psi (\boldsymbol{r};t=0)= e^{i \frac{v_x m x}{\hbar}} \Psi_0 (\mathbf{r})$  corresponds to an initial flux  $\mathbf{J}(\boldsymbol{r};t=0) = v_x  \rho (\boldsymbol{r};t=0)  \hat{x}$, with  $ \rho (\mathbf{r};t=0)=| \Psi_0(\boldsymbol{r})|^2$.
This allows us to define the local superfluid fraction as

\begin{align}
     f^{(\rm{NCTI})}_s &=\lim_{v_x\to0} \frac{\int_{r_{\rm min}}^{r_{\rm max}} dr\int d \mathbf{\hat{r}}\, \mathbf{J}(\mathbf{r};t \to \infty )  }{\int_{r_{\rm min}}^{r_{\rm max}} dr\int d \mathbf{\hat{r}}\, \mathbf{J}(\mathbf{r};0)}, 
     \label{eq:eq2}
\end{align}
where ($r_{\rm min},~r_{\rm max}$) are the radial boundaries of the transitional layer. The limit $v_x \to 0$ ensures the suppression of any extra excitation of the system.

\begin{figure}[t!]
\centering
\includegraphics[scale=0.42]{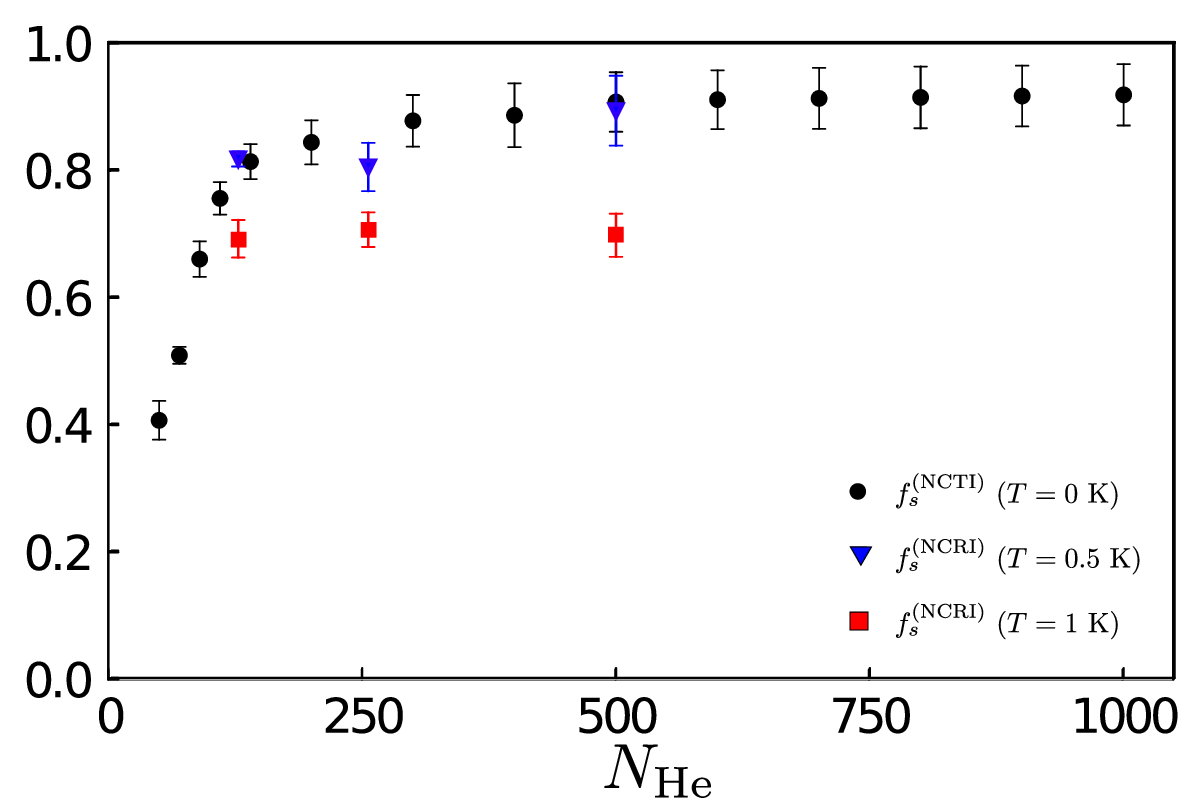}
\caption{The superfluid fraction for the transitional layer as a function of total number $N_{\rm{He}}$ of He atoms.  The non-classical translational inertia $f^{(\rm{NCTI})}_s$ \cref{eq:eq2} is calculated with mDFT (black dots) at zero temperature. Error bars indicate fluctuations of the mean value of $f^{\rm{NCTI}}_s$ for long propagation times.
The blue triangles and red squares show  PIMC calculations of the non-classical rotational inertia at $T=0.5~\rm{K}$ and $T=1~\rm{K}$.}
\label{fig:fig4}
\end{figure}

The superfluid fraction is equal to unity (zero) if all particles move with initial velocity $v_x$ (have come to rest) in the limit $t\to \infty$.
As seen in \cref{fig:fig4} (black dots), the superfluid fraction levels off at $ f^{(\rm{NCTI})}_s\sim0.9$ smaller than unity with increasing droplet size once the transitional layer has been completely filled (compare to \cref{fig:fig2}(d)), indicating its supersolid character. 
 
In addition, \cref{fig:fig4} shows the local superfluid fraction determined by non-classical rotational inertia (NCRI) with the PIMC approach by sampling the \textit{area estimator} \cite{likospre2001}.
Here, $ f^{(\rm{NCRI})}_s$ is obtained by averaging over the three orthogonal axes of rotation,  $ f^{(\rm{NCRI})}_s=(f_s^{(x)}+f_s^{(y)}+f_s^{(z)})/3$, denoted in \cref{fig:fig4} by blue triangles and  red squares at $T=0.5~\rm{K}$ and $T=1~\rm{K}$, respctively.
As expected, the superfluid fraction increases towards lower temperatures, and we observe that the mDFT superfluid fraction is in very good agreement with that measured with PIMC at $T=0.5$. The PIMC calculations confirm that the supersolid layer remains stable in the presence of thermal fluctuations. 
 
We can conclude that a $^4$He droplet doped by Na$^+$ has a highly structured interior surrounded by superfluid helium density. The interior consists of the snowball formed by
 two crystallized radial layers. A third, transitional layer is supersolid and builds the bridge to the surrounding superfluid density.
We have confirmed the density modulation-induced supersolidity of the transitional layer with Quantum Monte Carlo methods, revealing structured angular distributions and a substantial superfluid fraction less than unity characteristic for DMS. We found, even quantitatively, the same features with the hybrid approach we have developed by combining many-body G-TDH with mDFT. 
 
We expect that transitional supersolidity induced by density modulations can be found quite generally in settings, when geometry or other constraints restrict the symmetry breaking interaction to be comparable with  the particles' interaction in the superfluid. 
Obvious candidates are large He droplets doped with cationic dimers \cite{marinettiMicrosolvationCationicDimers2007},  with doubly ionized impurities, e.g. Ca$^{+2}$ \cite{zunzunegui2023observation}, or ions in Bose-Einstein condensates \cite{chowdhury2024ion}.
Our new hybrid method will allow the exploration of promising settings of supersolidity with large particle numbers in the future, which so far was impossible due to the computational 
cost of Quantum Monte Carlo methods.

\begin{acknowledgments}
This work was partly supported by the Austrian Science Fund (Grant
No. 10.55776/COE1) and the European Union (NextGenerationEU), and from the European Research Council through the ERC Synergy Grant SuperWave (Grant No. 101071882).
\end{acknowledgments}

%\bibliography{droplet}
%\makeatletter 
%\renewcommand{\thefigure}{S\@arabic\c@figure}
%\makeatother

\makeatletter 
\renewcommand{\theequation}{S\@arabic\c@equation}
\makeatother

\section{End Matter}
\subsection{A. The Gaussian time-dependent Hartree method} \label{sec:gtdh}

For the Hamiltonian shown in Eq. (1) in the main text,  we define $N_s$ Gaussian functions specified through the set of parameters: $ \boldsymbol{z}(\tau) = (\boldsymbol{z}_1(\tau), \boldsymbol{z}_2(\tau), .....,  \boldsymbol{z}_{N_s}(\tau))$ with $  \boldsymbol{z}_n(\tau)=(\gamma_n(\tau), \boldsymbol{r}_{0n} (\tau), \boldsymbol{p}_n(\tau)   ) $. 

The parameters $\boldsymbol{z}(\tau)$ are allowed to evolve in imaginary time until they converge to their equilibrium value. Here, $\boldsymbol{r}_{0n}$ represents the center of the n$^\mathrm{th}$ Gaussian, and $\boldsymbol{p}_n$ its momentum. The complex parameter $\gamma_n=\alpha_n+ \mathrm{i} \beta_n$ encodes the width (real part) and phase factor (imaginary part) associated to each Gaussian.

The snowball many-body ground state $\ket{\boldsymbol{z}(\tau)}$ is written as a direct product of single-particle Gaussians \cite{unn-tocQuantumDynamicsSolid2012}, neglecting all permutations,
\begin{align}
    \ket{\boldsymbol{z}(\tau)} &= a(\tau) \prod_{n=1}^{N_s} \ket{\boldsymbol{z}_n(\tau)},\\
    \braket{\boldsymbol{r}_n | \boldsymbol{z}_n} &= \left( \frac{2 \alpha_n (\tau)}{\pi} \right)^\frac{3}{4} e^{ -\gamma_n(\tau) |\boldsymbol{r}_n- \boldsymbol{r}_{0n}(\tau) |^2  +i \boldsymbol{p}_n(\tau) \cdot [\boldsymbol{r}_n- \boldsymbol{r}_{0n}(\tau) ] }\,. \label{eq:SnowballWavefunction}
\end{align}
This approximation is valid as long as the particles remain well localized, as is the case in the snowball regime.

Applying the Dirac-Frenkel-McLachlan variational principle \cite{raabDiracFrenkelMcLachlan2000a},  $ \braket{\delta \boldsymbol{z}(\tau) | \frac{\partial}{\partial \tau}  + \hat{H}  | \boldsymbol{z}(\tau) } =0$, we get equations of motion for the parameters in $\boldsymbol{z}(\tau)$:
\begin{align}
    &
    \begin{cases}
        \boldsymbol{\dot{r}}_{0n} &= \frac{\beta_n \boldsymbol{p}_n}{M_{\mathrm{He}} \alpha_n} -\frac{\nabla_{\boldsymbol{r}_{0n}} U }{2 \alpha_n}, \vspace{0.2cm} \\
        \boldsymbol{\dot{p}}_{n}  &= -\frac{2|\gamma_n|^2 \boldsymbol{p}_n}{M_{\mathrm{He}}\alpha_n}   - \frac{\beta_n}{\alpha_n} \nabla_{\boldsymbol{r}_{0n}} U, \vspace{0.2cm} \\
        \dot{\alpha}_n           &= -2 \frac{\alpha_n^2-\beta_n^2}{M_{\mathrm{He}}} + \frac{\nabla_{\boldsymbol{r}_{0n}}^2 U}{6}, \vspace{0.2cm} \\
        \dot{\beta}_n             &=-4\frac{\alpha_n \beta_n }{M_{\mathrm{He}}},
    \end{cases}  \label{eq:SnowballEOM}\\
     U(\boldsymbol{z}(\tau)) &=\braket{ \boldsymbol{z}(\tau) | \sum_{n=1}^{N_s} \hat{V}_{In}  + \sum_{m<n} \hat{V}_{nm} | \boldsymbol{z}(\tau) }. \label{eq:SnowballPotential}
\end{align}

Here, $\hat{V}_{In}$ represents the interaction potential of the ion of the $n-$th helium atom and $\hat{V}_{nm}$ is the interaction between atoms $n$ and $m$.

In order to obtain an analytical expression for \cref{eq:SnowballPotential}, and consequently for the equations of motion \cref{eq:SnowballEOM} we represent both the  He$-$He and  He$-$Na$^+$ interaction potentials, as a sum of three Gaussian functions ($\sum_{p=1}^{3} g_p e^{-b_p r^2}$). With this approximation, the strength of the potential is softened for short distances (in the repulsive region), also avoiding the singularity at the origin. The potential parameters are shown in \cref{tab:gaussian_interaction_potentials}.

%=======================================
%   Gaussian Potential Patameters Table
%=======================================
\begin{table}[h]
    \centering
	\setlength{\arrayrulewidth}{0.2mm} % Thicker lines
  \begin{tabular}{l S[table-format=4.9] S[table-format=4.9]} 
  %  \begin{tabular}{lll} 
		\hline\hline
		Parameters & {\centering  He$- $He} & { He$-$Na$^+$} \\
		\hline
		$g_1 [E_{\mathrm{H}}]$ & 0.059843  & 0.884099 \\
		\hline
		%\rowcolor{gray!30} % Sub-header color
		$g_2 [E_{\mathrm{H}}]$ & -0.000048 & -0.000655\\
		\hline
		$g_3 [E_{\mathrm{H}}]$ & -0.000751 &  -0.007144  \\
		\hline
		%\rowcolor{gray!30}
		$b_1 [a_0^{-2}]$ & 0.285038 & 0.410539  \\
		\hline
		$b_2 [a_0^{-2}]$ & 0.033862 &  0.022514 \\
		\hline
		%\rowcolor{gray!30}
		$b_3 [a_0^{-2}]$ & 0.111443 & 0.088281  \\
		\hline\hline
	\end{tabular}
	
	\caption{Parameters for modeling the  He$-$He and  He$-$Na$^+$ interaction potentials. These parameters were computed by fitting the sum of three gaussians to the potentials taken from \cite{ahlrichsInteractionPotentialsAlkali1988,ancilottoDensityFunctionalTheory2017}.}
	\label{tab:gaussian_interaction_potentials}
\end{table}
%============================================
%============================================

The wavefunction \cref{eq:SnowballWavefunction} describes a localized state if all coefficients $\alpha_n$ are positive; taking this into account, we note that $\beta_n=0$ is an attractor point of the equation for $\dot{\beta}_n$ in \cref{eq:SnowballEOM}. Hence, regardless of the choice of initial values, $\beta_n$ tends to zero. On the other hand, with initial momentum  zero (real-valued ground state), the momentum remains zero throughout the propagation.
Therefore, the set of equations \cref{eq:SnowballEOM}  reduces to the coupled system for $\boldsymbol{r}_{0n}$ and $\alpha_n$ and the state of atom $n$ can be represented as $\ket{\boldsymbol{z}_n(\tau)} \equiv \ket{\boldsymbol{r}_{0n}(\tau); \alpha_n(\tau)}$.

The parameter space of initial positions $\boldsymbol{r}_{0n}(0)$ and widths $\alpha_{n}(0)$ should be explored as much as possible in order to avoid convergence to a local minimum of the potential. In practice, we solve the set of equations \cref{eq:SnowballEOM} for a few thousand different initial conditions, and the least energetic configuration is chosen as the ground state of the system. 

%+++++++++++++++++++++++++++++++++++++++++++++++++++++++++++++
\subsection{B. Modified density functional theory for ion doped He droplets}
\label{sec:mdft}

The  delocalized ($N_d$) atoms beyond the snowball regime are described by the \textit{Order Parametrer} $\Psi(\boldsymbol{r})$, such that the single particle density is $ \rho(\boldsymbol{r})=\Psi^*(\boldsymbol{r}) \Psi(\boldsymbol{r})$. The total energy of the system is a functional of the single particle density (therefore of $\Psi$ and $\Psi^*$). We use the Orsay-Trento functional $E_\mathrm{OT}[\Psi,\Psi^*]$, whose parameters have been calibrated to give results in agreement with the experimental data \cite{dalfovoStructuralDynamicalProperties1995}.
In this functional the total energy of the system is computed by considering the contribution of: a) the kinetic energy of a system of non-interacting particles, together with a correction term, b) the  He$- $He interaction at large distances (Lennard-Jones potential truncated at short distances), and c) short-range correlation effects.

The $N_d$ atoms of the liquid evolve under the influence of both, the ion and the snowball external potentials. 
To obtain the ground-state solution, we minimize the modified energy functional $E[\Psi,\Psi^*]$ with respect to $\Psi^*$.
This leads to a Schr\"odinger-like equation with the density-dependent 
potential $V_{\rm mDFT}[\rho]$, 
\begin{align}
    \mu \Psi(\boldsymbol{r}) &= - \frac{\nabla^2 \Psi (\boldsymbol{r})}{2M_\mathrm{He}}  + V_\mathrm{mDFT}[\rho]\Psi(\boldsymbol{r})\equiv  \mathbb{H}[\rho] \Psi(\boldsymbol{r})
    \label{eq:mDFT}
\end{align}
 under the normalization constraint $\int d^3 \boldsymbol{r} \rho(\boldsymbol{r})=N_d$.

Assuming that $\Psi(\boldsymbol{r},\tau)= \Psi(\boldsymbol{r}) e^{-\tau \mu}$, then the ground state of \cref{eq:mDFT} is obtained by propagation in imaginary time according to
\begin{align}
    -\frac{\partial \Psi(\boldsymbol{r},\tau)}{\partial \tau} &= \mathbb{H}[\rho]  \Psi(\boldsymbol{r},\tau).\label{eq:DFT_EOM}
\end{align}
The equations \cref{eq:SnowballEOM} and \cref{eq:DFT_EOM} are solved self-consistently thereby taking  into account not only the influence of the snowball on the residual droplet, but also the impact of the droplet on the snowball.
The algorithm proceeds as follows: Once the droplet density is obtained via \cref{eq:DFT_EOM}, we propagate \cref{eq:SnowballEOM}  again including the corresponding density into \cref{eq:SnowballPotential}.
This procedure is repeated until the snowball configuration and density profile of the droplet are converged.

\end{document}